\documentclass[longauth]{aa}  %bibnumber
\usepackage{color}
\usepackage[colorlinks,breaklinks]{hyperref}
\hypersetup{linkcolor=blue,citecolor=blue,filecolor=black,urlcolor=blue}
\usepackage{amssymb}

\usepackage{xspace}

\usepackage{graphicx}
\usepackage{txfonts}
\usepackage{orcidlink}

\usepackage[switch]{lineno}
\begin{document}

   \title{ZTF SN~Ia DR2: Study of Type Ia supernova light-curve fits}
   \author{
    Rigault, M.\inst{\ref{ip2i}}\fnmsep\thanks{\texttt{m.rigault@ip2i.in2p3.fr}} \orcidlink{0000-0002-8121-2560},
    Smith, M.\inst{\ref{ip2i},\ref{lancaster}}\orcidlink{0000-0002-3321-1432},
    % -- %
    Regnault, N.\inst{\ref{lpnhe}},\orcidlink{0000-0001-7029-7901},
    Kenworthy, W. D.\inst{\ref{okc}}\orcidlink{0000-0002-5153-5983},
    Maguire, K.\inst{\ref{dublin}}\orcidlink{0000-0002-9770-3508},
    Goobar, A.\inst{\ref{okc}},\orcidlink{0000-0002-4163-4996},
    Dimitriadis, G. \inst{\ref{dublin}}\orcidlink{0000-0001-9494-179X},   
    Johansson, J.\inst{\ref{okc}}\orcidlink{0000-0001-5975-290X}, 
    % -- %
    Amenouche, M.\inst{\ref{canada}}\orcidlink{0009-0006-7454-3579},
    Aubert, M.\inst{\ref{clermont}},
    Barjou-Delayre, C.\inst{\ref{clermont}},
    Bellm, E.~C.\inst{\ref{dirac}}\orcidlink{0000-0001-8018-5348},
    Burgaz, U.\inst{\ref{dublin}}\orcidlink{0000-0003-0126-3999},
    Carreres, B.\inst{\ref{marseille},\ref{duke}}\orcidlink{0000-0002-7234-844X},
    Copin, Y.\inst{\ref{ip2i}}\orcidlink{0000-0002-5317-7518},
    Deckers, M.\inst{\ref{dublin}}\orcidlink{0000-0001-8857-9843},
    de~Jaeger, T.\inst{\ref{lpnhe}}\orcidlink{0000-0001-6069-1139},
    Dhawan, S.\inst{\ref{cambridge}}\orcidlink{0000-0002-2376-6979},
    Feinstein, F.\inst{\ref{marseille}}\orcidlink{0000-0001-5548-3466},
    Fouchez, D.\inst{\ref{marseille}}\orcidlink{0000-0002-7496-3796},
    Galbany, L.\inst{\ref{barcelona1},\ref{barcelona2}}\orcidlink{0000-0002-1296-6887},
    Ginolin, M.\inst{\ref{ip2i}} \orcidlink{0009-0004-5311-9301},
    Graham, M.~J.\inst{\ref{caltecphysics}}\orcidlink{0000-0002-3168-0139},
    Kim, Y.-L.\inst{\ref{lancaster}}\orcidlink{0000-0002-1031-0796},
    Kowalski, M.\inst{\ref{desy},\ref{humbolt}}\orcidlink{0000-0001-8594-8666},
    Kuhn, D.\inst{\ref{lpnhe}}\orcidlink{0009-0005-8110-397X}, 
    Kulkarni,\ S.~R.\inst{\ref{caltecphysics}}\orcidlink{0000-0001-5390-8563},
    M\"uller-Bravo, T. E., \inst{\ref{barcelona1},\ref{barcelona2}}\orcidlink{0000-0003-3939-7167},
    Nordin, J.\inst{\ref{humbolt}}\orcidlink{0000-0001-8342-6274},
    Popovic, B.\inst{\ref{ip2i}}\orcidlink{0000-0002-8012-6978},
    Purdum, J.\inst{\ref{caltecoptical}}\orcidlink{0000-0003-1227-3738},
    Rosnet, P.\inst{\ref{clermont}},
    Rosselli, D.\inst{\ref{marseille}}\orcidlink{0000-0001-6839-1421},
    Racine, B.\inst{\ref{marseille}}\orcidlink{0000-0001-8861-3052},
    Ruppin, F.\inst{\ref{ip2i}\orcidlink{0000-0002-0955-8954}},
    Sollerman, J.\inst{\ref{okc2}}\orcidlink{0000-0003-1546-6615},
    Terwel, J. H.\inst{\ref{dublin},\ref{not}}\orcidlink{0000-0001-9834-3439},
    Townsend, A.\inst{\ref{humbolt}}\orcidlink{0000-0001-6343-3362}
    }
    
   \institute{
%   Lyon
   Universite Claude Bernard Lyon 1, CNRS, IP2I Lyon / IN2P3, IMR 5822, F-69622 Villeurbanne, France
   \label{ip2i}
   \and
%   Lancaster
    Department of Physics, Lancaster University, Lancs LA1 4YB, UK \label{lancaster}
   \and
%   LPNHE
   Sorbonne Université, CNRS/IN2P3, LPNHE, F-75005, Paris, France
   \label{lpnhe}
   \and   
%   OKC
   The Oskar Klein Centre, Department of Physics, AlbaNova, SE-106 91 Stockholm, Sweden
   \label{okc}
   \and
%   Dublin
   School of Physics, Trinity College Dublin, College Green, Dublin 2, Ireland
   \label{dublin}
   \and
%   Canada   
   National Research Council of Canada, Herzberg Astronomy \& Astrophysics Research Centre, 5071 West Saanich Road, Victoria, BC V9E 2E7, Canada
   \label{canada}
   \and
%   Clermont
    Université Clermont Auvergne, CNRS/IN2P3, LPCA, F-63000 Clermont-Ferrand, France
    \label{clermont} 
    \and
%   Dirac
   DIRAC Institute, Department of Astronomy, University of Washington, 3910 15th Avenue NE, Seattle, WA 98195, USA
   \label{dirac}
   \and
%   Marseille   
   Aix Marseille Université, CNRS/IN2P3, CPPM, Marseille, France
    \label{marseille}
    \and    
%   Duke  
   Department of Physics, Duke University Durham, NC 27708, USA
   \label{duke}
   \and   
%   Cambridge   
   Institute of Astronomy and Kavli Institute for Cosmology, University of Cambridge, Madingley Road, Cambridge CB3 0HA, UK
   \label{cambridge}
   \and   
%   ICE Barcelona   
   Institute of Space Sciences (ICE-CSIC), Campus UAB, Carrer de Can Magrans, s/n, E-08193 Barcelona, Spain.
   \label{barcelona1}
   \and
%   IEEC Barcelona   
   Institut d'Estudis Espacials de Catalunya (IEEC), 08860 Castelldefels (Barcelona), Spain
   \label{barcelona2}
   \and   
%   Caltec Physics    
    Division of Physics, Mathematics \& Astronomy, California Institute of Technology, Pasadena, CA 91125, USA
    \label{caltecphysics}
   \and
%   DESI   
   Deutsches Elektronen-Synchrotron DESY, Platanenallee 6, 15738 Zeuthen, Germany
   \label{desy}
   \and   
%   Humbolt
   Institut für Physik, Humboldt-Universität zu Berlin, Newtonstr. 15, 12489 Berlin, Germany
   \label{humbolt}
   \and   
%   CalTech Optical
   Caltech Optical Observatories, California Institute of Technology, Pasadena, CA 91125
    \label{caltecoptical}
    \and
%   OKC2
   The Oskar Klein Centre, Department of Astronomy, Stockholm University, AlbaNova, SE-106 91 Stockholm, Sweden
    \label{okc2}
    \and
%   NOT   
   Nordic Optical Telescope, Rambla José Ana Fernández Pérez 7, ES-38711 Breña Baja, Spain
   \label{not}
%  \and    
    }
   \date{}

\titlerunning{ZTF SN Ia DR2: Light-curve residuals}
\authorrunning{M. Rigault et al.}
 \abstract{Type Ia supernova (SN~Ia) cosmology relies on the estimation of light-curve parameters to derive precision distances, which are used to infer cosmological parameters such as $H_0$, $\Omega_M$, $\Omega_\Lambda$, and $w$. The empirical SALT2 light-curve modeling that relies on only two parameters, a stretch $x_1$ and a color $c$, has been used by the community for almost two decades.
 %}
 %{
 We study the ability of the SALT2 model to fit the nearly 3000 cosmology-grade SN~Ia light curves from  the second release of the \textit{Zwicky} Transient Facility (ZTF) cosmology science working group.
 %}
%{
 While the ZTF data were not used to train SALT2, the algorithm models the ZTF SN~Ia optical light curves remarkably well, except for light-curve points prior to $-10$~d from maximum, where the training critically lacks data.
%}
 %{
 We find that the light-curve fitting is robust against the considered choice of phase range, but we show that the $[-10;+40]$~d range is optimal in terms of statistics and accuracy.
 %The lack of $i-$band data for a large fraction of the ZTF sample does not affect the accuracy of the SALT2 parameter estimation.
 %}
 %{
 We do not detect any significant features in the light-curve fit residuals that could be connected to the host environment. Potential systematic uncertainties associated tp population differences related to the SN~Ia host properties might thus not be accountable for by the inclusion of addition of light-curve parameters. However, a small but significant inconsistency between residuals of blue and red SN~Ia strongly suggests the existence of a phase-dependent color term, with potential implications for the use of SNe~Ia in precision cosmology. We thus encourage further work in this area to explore this possibility, and we emphasize that SN~Ia cosmology must include a SALT2 retraining to accurately model the light curves and avoid biasing the derivation of cosmological parameters.
 } 
% 5 {} token are mandatory


   \maketitle

\section{Introduction}
\label{sec:intro}

Type Ia supernovae (SNe~Ia) are standardizable cosmological candles. With a limited set of parameters that are usually derived from light-curve observations, relative distances can be derived to a precision of about 7\% up to a few gigaparsec thanks to the remarkable SN Ia brightness at optical wavelengths ($<-19$ mag). These characteristics make SNe~Ia a key cosmological tool for probing the expansion history of our Universe, which led to the discovery of the acceleration of its expansion (\citealt{perlmutter1999,riess1998}; see \citealt{goobar2011} for a review). SNe Ia also play a central role in the derivation of the Hubble-Lemaître constant, the value of which is currently highly debated \citep{planck2020, freedman2019,riess2022}.

For the past 25 years, two light-curve parameters have been used to standardize SNe~Ia \citep{tripp1998}. The standardization exploits two empirical relations: redder SNe~Ia are fainter \citep[color yields to the redder-fainter relation;][]{riess1996}, and more slowly evolving SNe~Ia are brighter \cite[stretch yields to the brighter-slower relation;][]{phillips1993}. When we account for these two linear relations, the natural SN~Ia Hubble diagram dispersion is reduced from 0.40~mag to 0.15~mag \citep[e.g., recent SN~Ia compilations][]{betoule2014, scolnic2018,brout2022}. Half of this variance can be explained by measurement or modeling errors. This leaves $\sim0.10$~mag unexplained, which is usually referred to as the intrinsic scatter. 

The light-curve parameters of stretch and color are derived using light-curve fitter algorithms. The spectral adaptive light curve template \citep[SALT2,][]{guy2007,guy2010,betoule2014} has been the community standard for the past decade. This data-driven model is based on a principal-component analysis (PCA) with an exponential phase-independent color law. The PCA and color-law parameters are trained on calibrated light curves and spectra to build a spectro-temporal model. The first component $M_0$ (eigenvalue $x_0$) corresponds to the spectral template for an average SN~Ia, while the second $M_1$ (eigenvalue $x_1$) corresponds to the phase and amplitude deformation of the $M_0$ component. The third component has been shown to contain insignificant information \citep[e.g.,][]{guy2007}. 
The amplitude of the multiplicative color correction is called $c$. For SALT2, $x_1$ corresponds to the stretch, $x_0$ to the amplitude (flux), and $c$ to the color. The SN light-curve shape is therefore captured using only the two parameters $x_1$ and $c$. The most frequently used SALT2 training is that from \cite{betoule2014} and is usually referred to as SALT2.4. \cite{taylor2021} provided a recent retraining of SALT2.4 with additional data. \cite{kenworthy2021} redesigned the original SALT software and trained it with a recent SN~Ia cosmological compilation \citep{scolnic2018}. It was released as SALT3. 
In addition to the SALT models, the community has developed alternative ponent light-curve models such as MLCS \citep{jha2007}, SNooPy \citep{burns2011}, BayesSN \citep{mandel2022, grayling2024}, and PISCOLA \citep{mullerbravo2022}. They differ from SALT in that they incorporate additional physics, most notably in the treatment of dust reddening.

The stretch- and color-standardized SN~Ia magnitudes have been shown to significantly depend on their host-environments, however. SNe Ia from massive host galaxies or redder UV-optical color environments are brighter (after standardization) than those from lower-mass (bluer) hosts \citep[e.g.,][]{sullivan2010, roman2018, rigault2015, rigault2020, briday2022}. This is usually refereed to as the “mass-step” because it was first observed using global host stellar mass tracers \citep{kelly2010,sullivan2010}. The actual origin of this effect is highly debated. It might either be caused by differences in progenitor age, "prompt versus delayed" \citep{rigault2013,rigault2020,roman2018,briday2022}, or it might originate from a variation in the dust color law in the different environments \citep{brout2021,popovic2021}. Both might be true, or both might be related aspects of a one underlying parameter \citep{wiseman2022,kelsey2023}. In any case, understanding the origin of these correlations is important because any differences in their evolution with redshift is likely to impact the inferred cosmological parameters.

An ideal solution would be to directly find signs of this possible SN-host correlation in the SN~Ia light curves. If this were the case, it might empirically corrected for, as is done for the stretch and color, in the hope of capturing the entire variance in the SN~Ia magnitude.
In this context, alternative SN~Ia models have emerged. \cite{saunders2018} and \cite{leget2020} used spectroscopic information, such as silicon velocities and equivalent widths at maximum light, instead of the stretch to standardize SNe~Ia. 
They were able to reach a dispersion of $\sim0.15$~mag with five or more parameters, but did not see a significant reduction in the Hubble diagram scatter. Their results are still affected by significant environmental biases. 
A promising way forward is the twin technique. Introduced by \cite{fakhouri2015} and based on the spectrophotometric nearby supernova factory sample (SNfactory; \citealt{aldering2002}), the idea is that two SNe~Ia with a nearly identical spectral time-series probably have an intrinsically similar absolute brightness. Advances in machine-learning methods enabled \cite{boone2021a} and \cite{boone2021b} to simplify the twinning down to three intrinsic parameters that were estimated near maximum light. The authors demonstrated that twinning, based on spectroscopy rather than light-curve photometry, provides a scatter as low as $\sim0.07$~mag, which leads to a strong reduction of the astrophysical biases, but not to a full cancellation. However, acquiring spectrophotometry is difficult, and this technique has not been demonstrated to work beyond the SNfactory sample. Future space missions such as the \emph{Nancy Roman} Space Telescope might be able to acquire data like this through slitless spectroscopy at high redshifts \citep{rose2021,rubin2022}.

Twinning has nonetheless demonstrated that additional information in the SN data is available that is not yet captured by classical photometric light-curve fitters. The question is whether this information can be identified in the phase-evolution of multiband photometry, or if this is solely accessible with spectroscopic data. 

In this paper, we study the SN~Ia light-curve residuals of the second data-release of the \textit{Zwicky} Transient Facility \citep[ZTF;][]{bellm2019, graham2019} cosmology science working group (DR2; \citealt{dr2rigault}).
This release contains SN~Ia light curves with a daily cadence and unprecedented phase coverage. This dataset is consequently particularly well suited to probing deviations from the current light-curve templates and to unveiling new features. Increased diversity has been observed at both early ($\sim15$~d prior to maximum light) and late time ($>30$~d after maximum light), which suggests that these phases contain key information about SN~Ia physics and its diversity \citep[e.g.,][]{dimitriadis2019,deckers2022,deckers2023}. With training samples dominated by higher-redshift data around maximum light \citep{taylor2021}, this diversity is likely not encoded in the SALT2 model. The unique phase coverage of ZTF SN~Ia DR2 allows us to test how well this model describes the SN~Ia population as a function of phase, and hence, to deduce the exact phase range that should be used for cosmological inference.

We start the paper with a short introduction of the DR2 sample in Sect.~\ref{sec:dr2overview}. We then present the SN~Ia SALT2 light-curve residuals in Sect.~\ref{sec:lcfitaccuracy}. The observed light-curve residual deviations are studied in detail in Sect.~\ref{sec:lcres}, notably, variations as a function of light-curve and host parameters. The robustness of the light-curve parameter estimation released as part of DR2 is presented in Sect.~\ref{sec:lcparam}. We conclude in Sect.~\ref{sec:conclusion}.

Throughout the paper, the phases are in days, in the rest frame and set to 0 at $t_0$, the estimated date of maximum light, except when explicitly indicated otherwise.

%
%
%  SECTION 2: ZTF Cosmo DR2 Review
%
%
\section{ZTF cosmology DR2}
\label{sec:dr2overview}

The second data release of the ZTF Cosmology Science working group is presented in \cite{dr2rigault}. It contains 3628 spectroscopically confirmed SNe Ia with a median redshift of $z=0.08$. The light curves have a typical three-day cadence in ztf:g ($g$) and ztf:r ($r$) and a five-day cadence in ztf:i ($i$). Of these, 2625 are nonpeculiar SNe~Ia that passed the basic cuts (a good sampling and reasonable light-curve fit parameters; see \citealt{dr2rigault}).
In the $-10$ to $+40$ rest-frame phase range, all of these 2625 have at least seven detections, two of which must be before (pre), and two others must be after (post) maximum light\footnote{all but one have both $g$ and $r$ data. This one target without \textit{g} and \textit{r} data has $g$ and $i$, but no $r$}. In addition, all targets must have detections in at least two bands. Nearly half of our targets (46\%) have $i$-band data.

The unique ZTF cadence typically allowed a first detection two weeks prior to maximum light, and 80\% of our 2576 SNe~Ia had at least one detection prior to $-10$ d. In the $[-20, +50]$~d phase range, our targets have an average of 20 detections pre maximum light and 50 detections post maximum light. Eighty-five percent of the targets have at least one detection after $+30$~d. This unprecedented sample size makes the ZTF SN~Ia DR2 dataset a unique opportunity for studying SN~Ia light curves in detail. 

Except when indicated otherwise, the default light-curve fitter algorithm was \texttt{SALT2} \citep{guy2010}, as available in \texttt{sncosmo} \citep{sncosmo2023}. 
By default, we used the SALT2 version T21 from \cite{taylor2021}, which corresponds to a small calibration update of the reference version 2.4 \citep{betoule2014}. As explained in Sect.~\ref{sec:lcfitaccuracy}, the default fitted phase range is $\phi\in[-10,+40]$ d. As expected from \citet{taylor2021}, we did not notice significant differences with the SALT2.4 surfaces \citep{betoule2014} instead of the T21 surfaces. Because the latter benefit from a larger training dataset, however, we decided to favor the new T21 retrained surfaces (see \citealt{dr2rigault} for details of the light-curve extraction and fitting procedure).

%
%
%  SECTION 3: LC Residuals big picture
%
%
\section{Generic overview of SN~Ia light-curve fitting.}
\label{sec:lcfitaccuracy}

\begin{figure*}
  \centering
  \includegraphics[width=\linewidth]{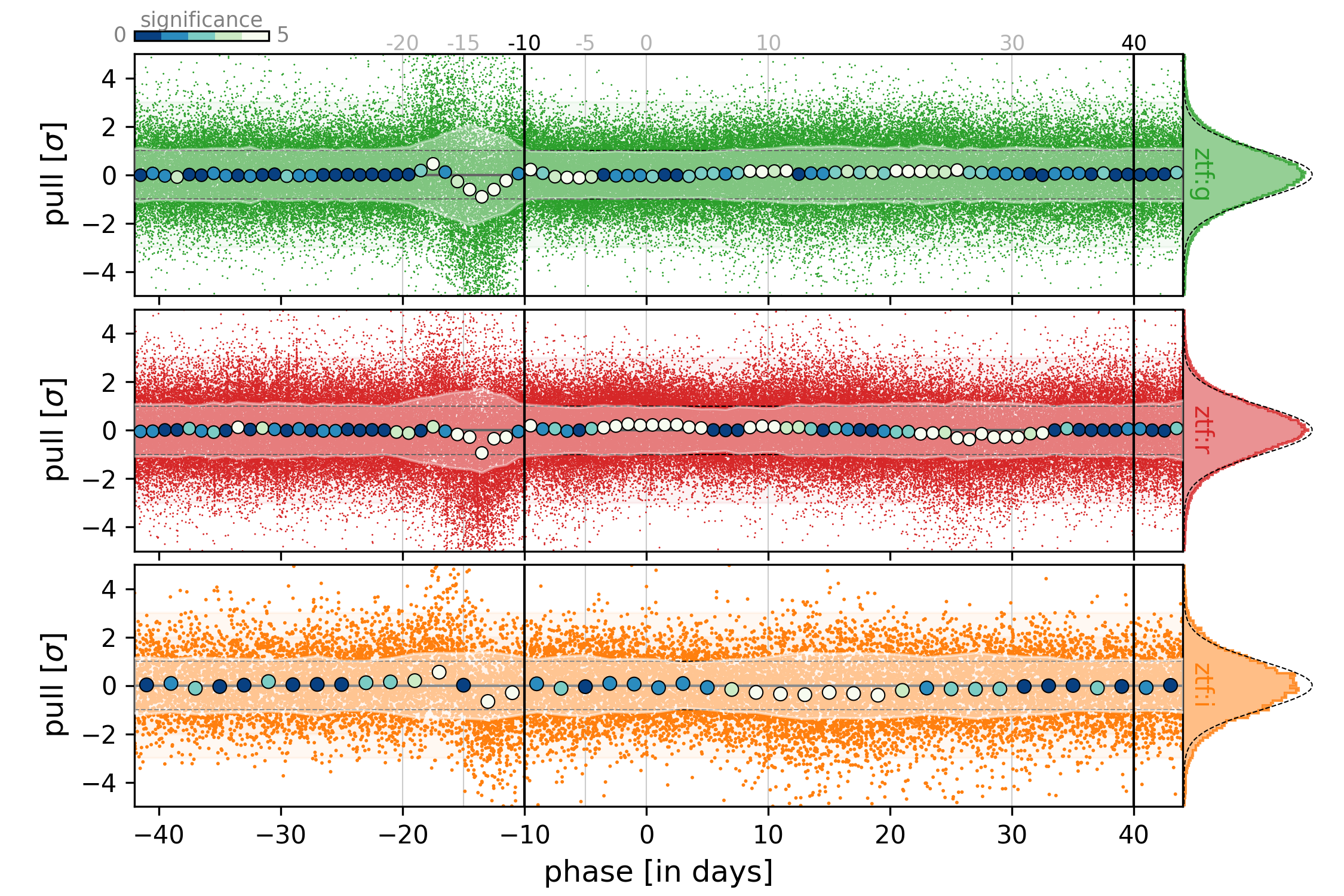} % .pdf available
  \caption{
    Best-fit light-curve residual in units of errors (i.e., the pull) as a function of rest-frame phase. From \emph{top} to \emph{bottom}, we show the $g$- (green), $r$- (red), and $i$-band (orange).  The errors include the band-dependent error floor (see Sect.~\ref{sec:lcfitaccuracy}). 
    In each panel, small colored markers represent individual data points (pull), and blue-to-white makers show the median pull per daily bin (2-d for $i$ band). These markers are colored by the significance of their deviation from zero. White marker are nonzero at least at the $5\sigma$ level (see the top left significance color bar). 
    The running white bands centered on zero show the bins for nMAD. 
    If the light-curve model were to perfectly represent our data, and if our errors were Gaussian and  correctly estimated, the pull should be centered on zero with a scatter of one. The thin dashed horizontal lines show the $±1\sigma$ for reference. The histograms in the \emph{right panel} show the phase-marginalized distribution per band. An $\mathcal{N}(0,1)$ distribution is also displayed as a thin dashed line for reference. The vertical black lines show the phase range used to fit the light-curve data $\text{phase} \in [-10, +40]$ d. This figure contains 2625 SNe~Ia.
  }
  \label{fig:lcresidual_main}
\end{figure*}

We summarize the light-curve residuals for the 2625 SNe~Ia from the DR2 sample in Fig.~\ref{fig:lcresidual_main}. The light curves were modeled as the best SALT2.4 (T21) light-curve template fit in the $\phi \in [-10, +40]$~d rest-frame phase range on all three ZTF bands (see Sect.~\ref{sec:lcres}).
This figure presents the normalized light-curve residuals, that is, the $\text{pull} \equiv \left(\text{data}-\text{model}\right) / \text{error}$) where errors are a quadratic sum of the actual statistical measurement errors, a model error (subdominant), and a filter-dependent error floor, estimated as a fraction of the observed flux. This error floor was estimated to obtain a pull normalized median absolute deviation (nMAD) of one for the fit phase range and corresponds to unaccounted-for errors related to the forced-photometry transient signal extraction. We find that $g$, $r$, and $i$ SN photometry need a 2.5\%, 3.5\%, and 6\% error floor, respectively. This agrees well with the simulations presented by \cite{dr2amenouche} as part of the ZTF SN~Ia DR2 release. 
This additional error floor typically corresponds to $\sim20\%$ of the total error in the $\phi \in [-10, +40]$~d phase range, and to up to $\sim35\%$ near maximum light.
As illustrated in Fig.~\ref{fig:lcresidual_main}, the full light-curve error leads to an expected pull scatter of $±1\sigma$ at all wavelengths (white bands). We emphasize that the add-on error floor has no effect prior $\phi=-20$~d as there is no SN flux yet at these phases. 
In the figure, the residuals are binned per phase (1~d for $g$ and $r$, and 2~d for $i$). We display the median (used as a robust mean) residual per bin and computed its error as the error on the mean ($\text{nMAD}/\sqrt{N-1}$), with the nMAD used as a robust STD. The significance shown in Fig.~\ref{fig:lcresidual_main} (and similar figures) corresponds to the deviation of the median from zero given its error.

The pre-explosion epoch ($\phi<-20$~d) in Fig.~\ref{fig:lcresidual_main} is also interesting for two reasons. First, the $\phi [-50, -20]$~d phase range was not considered during the forced-photometry baseline correction (see (Smith et al., in prep.). Therefore, this allowed us to test the quality of this procedure, which consisted (1) of shifting the light-curve fluxes to account for potential SN contamination or defects in the reference images, and (2) of scaling the measured flux errors to have a pull scatter of one when only noise is expected. Fig.~\ref{fig:lcresidual_main} shows that the residuals are centered close to zero with a scatter of one, as expected. Second, because there is no significant excess in the $\phi \in[-40, -20]$~d period, we can exclude the existence of systematic significant pre-explosion features.

At epochs following the explosion, the model shows many small- to medium-size deviations from what is expected, as shown in Fig.~\ref{fig:lcresidual_main}. These are discussed in detail in Sect.~\ref{sec:lcres}.

\section{Study of the light-curve residuals}
\label{sec:lcres}

 %presents deviations at the $5\sigma$ level (or more) at many phases. We define  characteristic epochs that are discussed in Sect.~\ref{sec:resepochs}. 
 
At first glance, it is generally remarkable that the empirical SALT2 model, which was trained on a much smaller dataset than this ZTF data release and made of SNe~Ia at higher redshift, performs so well. This is particularly true near maximum light ($\phi \in [-5,+20]$~d), where most of the flux information is extracted to derive cosmological distances. 
%Second, the density of $g$- and $r$-band data is so high in comparison to $i$-band, that the SALT2 color estimation is fully dominated by these two bands. Yet, the $i$ band residual is globally consistent with zero on average, which asserts the accuracy of the current relative band photometric calibration (see further discussion concerning the $i$ band in Sect.~\ref{sec:gri_vs_gr}).

We first discuss the residuals in characteristic epochs of the light curves in Sect.~\ref{sec:resepochs}.
We then investigate in Sect.~\ref{sec:phasefit} the impact of narrowing the fit phase range and discuss our selection of the  $\phi\in[-10,+40]$~d phase range for our baseline light-urve fits in ZTF SN~Ia DR2. We continue our investigation of the light-curve residuals in Sect.~\ref{sec:lcresvslcparam}, where we analyze how they vary as a function of SN parameters. In Sect.~\ref{sec:lcresvshost} we analyze how they vary as a function of SN environmental properties.

\subsection{Analysis of the main light-curve epochs.}
\label{sec:resepochs}

\begin{figure}
  \centering
  \includegraphics[width=\linewidth]{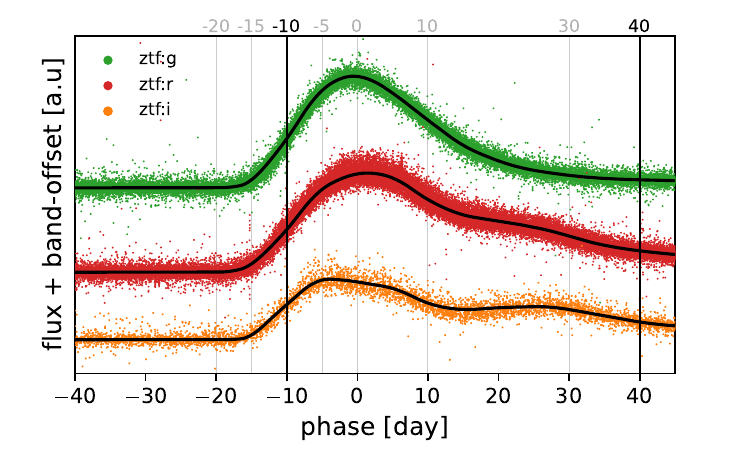} % .pdf available
  \caption{Example of Type Ia supernova light curves normalized by their SALT2 flux intensity parameter $x_0$. 
  For this illustration, only targets passing the following criteria are displayed (see Sect.~\ref{sec:resepochs}): $z\in[0.05, 0.09]$, $c\in[-0.1,0.1]$, and $x_1\in [-1,1]$ . In each panel, the black lines represent the SALT2 model at the median redshift ($z=0.07$), color ($c=0.01$), and stretch ($x_1=0.17$). 
  \emph{Bottom panels}: Zoom on specific time periods illustrated by a given band. From top to bottom, early phases ($g$), near maximum light ($r$), and the second peak period ($i$) are shown. White markers show the daily bin median fluxes (quarter of a day for the early phases).
  This figure contains 781 SNe~Ia.
  }
  \label{fig:lcexample}
\end{figure}

Figure~\ref{fig:lcexample} illustrates typical SN~Ia light curves and SALT2 models using ZTF SN~Ia DR2 data. The SN Ia data shown in this figure are normalized by their best-fit flux at maximum light ($x_0$), and they are corrected for Milky Way extinction \citep{Schlafly2011}. For the sake of visibility, the figure only contains targets within a redshift range of $z\in[0.05, 0.09]$, a color range of $c\in[-0.1,+0.1]$, and a stretch range of $x_1\in [-1,+1]$, so that all displayed SNe~Ia are comparable.
This figure illustrates three periods, which we discuss in detail below: (1) early phases, that is, the first days following the explosion ($\phi\in[-18,-10]$~d); (2) maximum light ($\phi\in[-5,+10]$~d); and (3) the second peak in the red to near-infrared bands ($\phi\in[+15,+35]$~d), which is a noticeable SN~Ia light-curve feature.

\subsubsection{Early phases: $\phi \in [-18,-10]$~d.}
\label{sec:resepochsearly}

Early SN~Ia epochs are well probed by the ZTF. This period is particularly important for progenitor studies because it contains imprints of the progenitor channel(s), and early light-curve excesses are particularly searched for \citep[see, e.g.,][and references therein]{deckers2022}, as is the rise time \citep[e.g.,][]{firth2015}. In turn, a deeper understanding of SN~Ia physics provides key information for SN~Ia cosmological analyses for understanding and mitigating the origin of astrophysical biases \citep[see, e.g.,][]{dr2senzel} and reference therein). 

However, this very early phase range currently lacks training data, and existing light-curve fitting algorithms mostly extrapolate below $\phi=-10$~d (see the data made available by \citealt{guy2010,betoule2014}, \citealt{taylor2021}, and \citealt{kenworthy2021}).
The strong deviations in this phase range, which was largely unexplored before the ZTF, are therefore expected. In Fig.~\ref{fig:lcresidual_main} and in the early phase panel of Fig.~\ref{fig:lcexample}, the model overestimates the data fluxes in the $\phi=-15$ to $-10$~d range. 
Figure~\ref{fig:lcresidual_main} also shows a flux excess  in the $\phi=-18$ to $-15$~d period, especially in $i$ and $g$. The SALT2 model does not prohibit the model flux from becoming negative, which is not physical. This is particularly true for low $x_1$ models. The early excess can thus be attributed to a modeling issue and not to a real physical effect.

In the $\phi=-15$ to $-10$~d period, the model overprediction suggests that the SN~Ia light-curve rise starts slightly later and rises slightly faster than predicted by the model. This is likely a regularization issue that penalizes fast evolution of the model eras that lack sufficient training. At $\phi=-10$~d, the SALT2 model converges toward the correct SN~Ia rise pace, likely because of sufficient training.
This is visible in Fig.~\ref{fig:lcexample}: The black line (model) clearly lies on top of the white marker (binned data).

\subsubsection{Near-maximum phases: $\phi \in [-5,+15]$~d.}
\label{sec:resepochsmax}

The few deviations in Fig.~\ref{fig:lcresidual_main} when the SALT2 model was applied to our multi-thousand SN~Ia dataset that never entered the model training are remarkable. The near-maximum epoch range indeed has the highest spectral and photometric training density. The observation that light-curve residuals are globally flat and consistent with zero is therefore expected. Nonetheless, this accuracy is reassuring for SN cosmology because this particular phase range, $\phi \in [-5,+15]$~d, is most important for determining the SN light-curve peak flux, stretch, and color, from which distances are ultimately derived.

A careful inspection of Fig.~\ref{fig:lcresidual_main} reveals a small but significant flux excess in $r$, however. The model slightly underestimates the observed flux in the week of maximum light. Hence, either the SALT color law (CL) is slightly inaccurate for this band ($\lambda \in [5650, 7250]\,\AA$), or the underlying spectral template model ($M_0$) is inaccurate in this phase range. This is further investigated in Sect.~\ref{sec:lcresvslcparam}, where we analyze the variation in light-curve residuals as a function of light-curve parameters.

\subsubsection{Second bump phases: $\phi \in [+15,+35]$~d.}

The second-peak period at $\phi \in [+15, +30]$~d corresponds to the arrival of heavy elements such as Fe II in the explosion ejecta (powered by the decay chain $^{56}\mathrm{Ni} \rightarrow{} ^{56}\mathrm{Co} \rightarrow{} ^{56}\mathrm{Fe}$), which, combined with a change in opacity, produces a flux increase in the red to near-infrared wavelengths \citep{kasen2006}. This period shows an interesting increase in the SN~Ia luminosity, which is thought to depend on progenitor and/or explosion mechanism properties \citep{dhawan2015}. The current SALT template model has no surfaces that would be particularly sensitive to this epoch, however \citep[e.g.,][]{kenworthy2021}. An extension of the model to further probe this unique SN~Ia feature, for instance, with an additional phase-dependent parameter (e.g., a second stretch) may reduce the SN~Ia scatter in the Hubble diagram or account for the observed correlations between the SN~Ia standardized magnitude and the properties of their environment \citep[see, e.g.,][and references therein]{folatelli2010, papadogiannakis2019, pessi2022}.

Fig.~\ref{fig:lcresidual_main} shows a small but long (nearly two weeks) model overestimation of the flux in the $i$ band. In the example shown in Fig.~\ref{fig:lcexample}, the data clearly dip below the model at $\phi=+15$~d.

This smoothness issue is again likely due to a limited number of rest-frame wavelength data redder than $7000\,\AA$ because the SALT2 training sample is usually made of higher-redshift targets \citep{betoule2014,taylor2021}. For cosmology, the inability to correctly model the phase evolution of SN~Ia light curves (the model is too smooth at $+15$ d, but also at $-15$ d; see Sect.~\ref{sec:resepochsearly}) could lead to an inaccurate estimation of the light-curve parameters ($x_0$, $c$, and $x_1$), and then in turn to errors in the derivation of distances. A better modeling could thus improve the overall light-curve fit. Furthermore, an improvement like this in the red to near-infrared bands would help photo-typing methods because this second peak feature is unique to SNe Ia \citep[e.g.,][]{kasen2006}. 

\subsection{Study of the phase-range fit}
\label{sec:phasefit}

When fitting a light curve, the desire is to include a phase range that is as large as possible to better constrain the model parameters. This is especially true for the stretch. Poorly modeled phase ranges will bias the parameter estimations, however. It is therefore expected that phases outside the range $\phi \in [-15, +45]$~d, which have nearly no training data \citep{betoule2014, taylor2021, kenworthy2021}, should not be included. 

\begin{figure}
    \centering
    \includegraphics[width=0.95\linewidth]{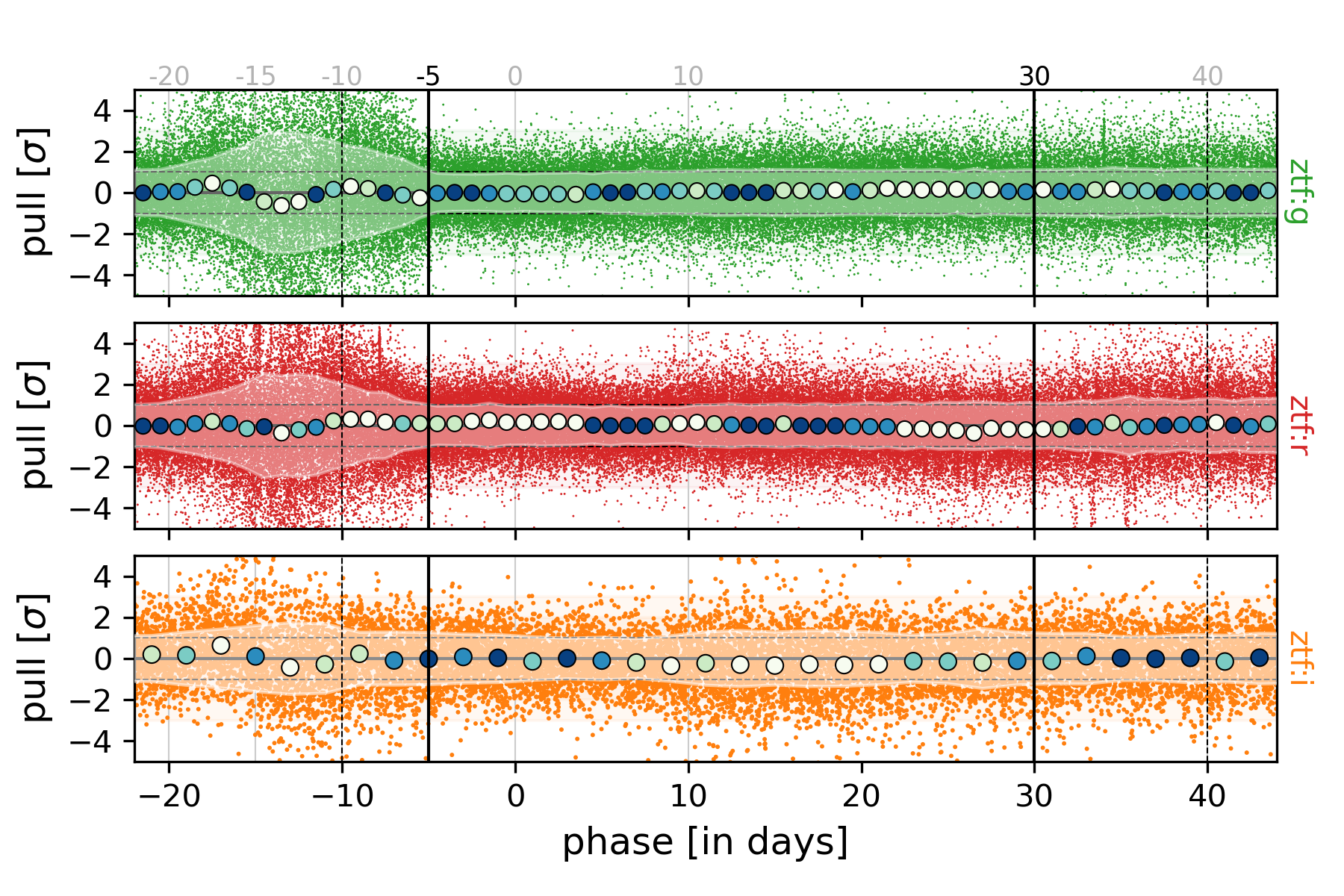} % .pdf available
    \caption{Similar to Fig.~\ref{fig:lcresidual_main}, 
    but with a fit phase range $\phi \in [-5, +30]$~d instead of the default $\phi \in [-10, +40]$~d phase range (vertical black lines).}
    \label{fig:lcresidual_phase}
\end{figure}

Figure~\ref{fig:lcresidual_phase} shows the light-curve residuals when light curves were fit using the $\phi \in [-5, +30]$~d phase range. 
The structure is very similar to that in Fig.~\ref{fig:lcresidual_main}, showing that the inclusion of  $\phi \in [-10, -5]$~d and $\phi \in [+30, +40]$~d phases does not bias the light-curve modeling: SN~Ia light curves are well modeled in the extrapolated $\phi=[-10,-5]$ and $[+30, +40]$~d phase ranges because the residuals are consistent with zeor in all three ZTF bands. 

Hence, to obtain the most accurate and precise light-curve fits, we considered the wider well-modeled phase range for the ZTF SN~Ia DR2 fits, and we thus advocate using $\phi=[-10, +40]$~d. This is consequently the rest-frame phase range used for the DR2 release (\citealt{dr2rigault}, Smith et al., in prep.). 

\subsection{Goodness of  fit as a function of light-curve parameters}
\label{sec:lcresvslcparam}

We present in Fig.~\ref{fig:lcresidual_cvar} and Fig.~\ref{fig:lcresidual_x1var} the light-curve residuals after splitting our SN~Ia sample according their light-curve color ($c$) and stretch ($x_1$), respectively. 
This enabled us to investigate the potential origin of the observed deviations. 
For instance, residual variations observed regardless of the light-curve parameters would point toward issues in the average template $M_0(\phi, \lambda)$, while color-dependent but phase-independent residuals would point toward 
a miscalibration of the SALT2 color law $CL(\lambda)$.
However, we highlight that SN Ia light-curve features not captured by the SALT2 model would likely lead to residual variation scatter across the entire light-curve phase range. By splitting the SNe along a parameter that might be related to the missing part of the model (e.g., observed host biases), this effect might therefore be unveiled.

We investigate the color dependences (Fig.~\ref{fig:lcresidual_cvar}, Sect.~\ref{sec:lcrescolordep}) first and the stretch dependences (Fig.~\ref{fig:lcresidual_x1var}, Sect.~\ref{sec:lcresx1dep}) second. 
The host dependences are presented in Sect.~\ref{sec:lcresvshost}.

\subsubsection{Color dependence}
\label{sec:lcrescolordep}

\begin{figure}
    \centering
    \includegraphics[width=\linewidth]{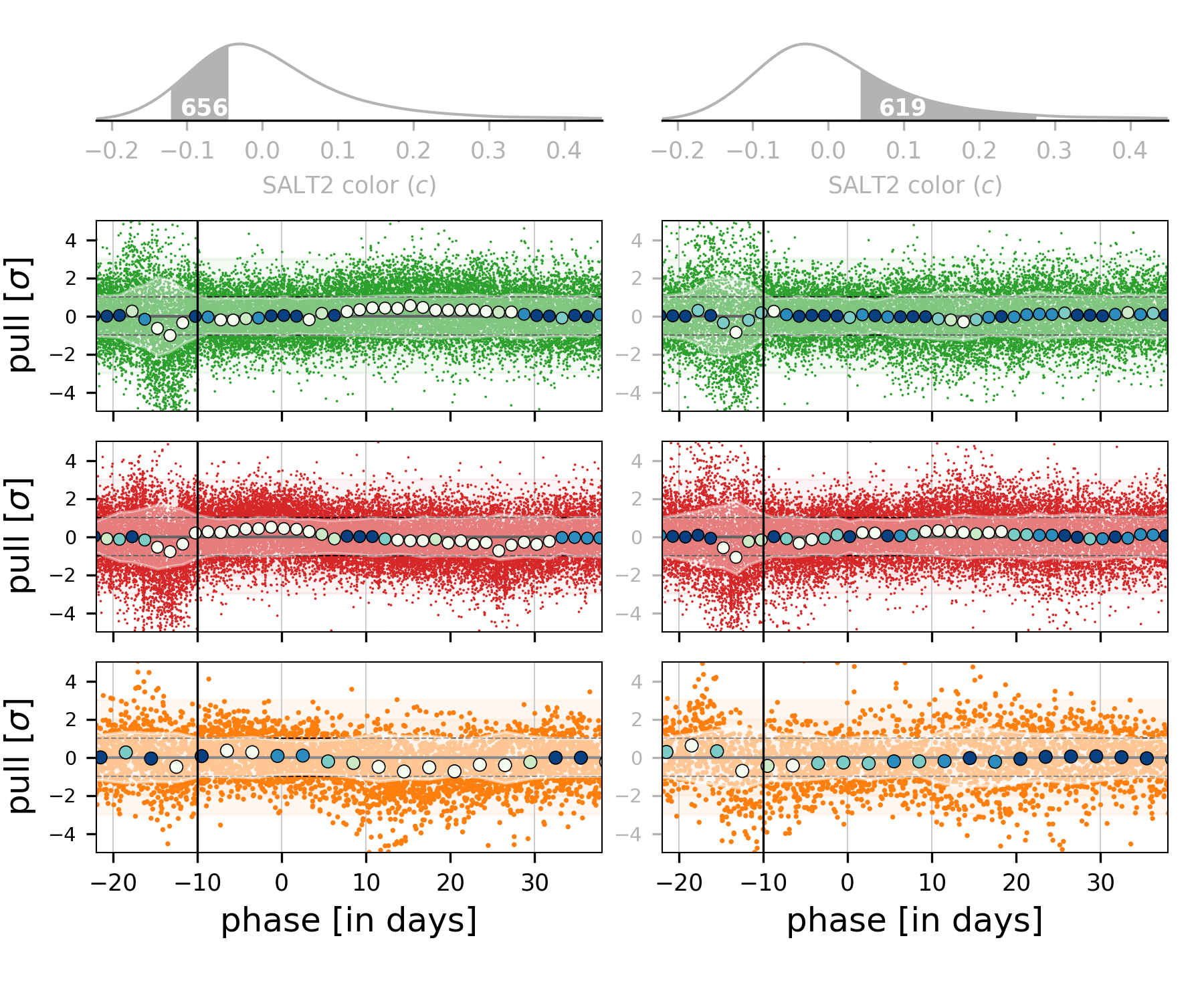} % .pdf available
    \caption{Similar to Fig.~\ref{fig:lcresidual_main}, 
    but split per light-curve color ($c$) to separately probe blue and red SNe~Ia. 
    For visibility, the phase bins are 1.5~d in $g$ and $r$ and 3~d in $i$.
    \emph{Left side:} $c\in[5\%, 30\%]$ percentile SNe~Ia. \emph{Right side:} $c\in[70\%, 95\%]$ percentile SNe~Ia. 
    \emph{Top:} SN~Ia color distributions highlighting the part considered on each side, indicating the corresponding sample size.
    }
    \label{fig:lcresidual_cvar}
\end{figure}

The SN~Ia color is thought to be a mixed contribution of intrinsic color variations (dominating at $c<0$) and external extinction caused by interstellar dust of the host (dominating at $c>0.2$; see, e.g., \citealt{brout2021,popovic2021} and references therein).
Figure~\ref{fig:lcresidual_cvar} separately shows the light-curve residuals of bluer and redder SNe~Ia.
Surprisingly, redder SNe~Ia ($c \in [0.05, 0.28]$, the 70\%-95\% $c$ range) appear to be globally better modeled by SALT than bluer ones ($c\in[-0.11, -0.03]$, 5\%-30\%) because they have weaker deviations. This is surprising because the SN~Ia light-curve modeling more strongly depends on the $CL(\lambda)$ when $|c|$ is high. Conversely, $CL(\lambda)$ has no effect for a $c=0$ target. 

When the blue SNe~Ia in Fig.~\ref{fig:lcresidual_cvar} are considered in detail, we note that before maximum light, the $r$- and $i$-band light curves are underestimated (negative pull), but the $g$-band light-curves are overestimated. This reverses at phase $\phi=+10$~d and converges toward zero only at phase $\phi=+30$~d.
This behavior is not detected for redder targets.

We recall that the SALT color law $CL(\lambda)$ is phase independent, and the first template order, $M_0(\phi,\lambda)$, does not depend on any light-curve parameter. In other words, the red and blue SN~Ia light curves are modeled by the same $M_0(\phi,\lambda)$ and are deformed by $c\times CL(\lambda)$, and this deformation is independent of the phase. It is thus intriguing to observe SN-color and phase-dependent residual variations. 
%an ill-modeling of $M_0(\phi,\lambda)$ could explain the phase-evolution observed in blue SNe~Ia, but a similar residual should also be seen in red SNe~Ia, which is not the case. Similarly, an ill-model of $CL(\lambda)$ could affect SN~Ia differently as a function of their color, but the residual effect should be phase-independent. 
In principle, the stretch and color corrections are independent. To ensure that what we report is not affected by an $x_1$-related issue, however, we repeated the same analysis considering $x_1\in[-0.5,+1.5]$ or $x_1\in[-2.0,-0.5]$ SNe~Ia only.  
We found exactly the same results when we compared blue and red SNe~Ia. Hence, the phase-dependent features visible in Fig.~\ref{fig:lcresidual_cvar} are not related to potential $M_1(\phi,\lambda)$ template issues.

It is consequently very likely that an extension of the SALT2 model is required to capture the observed phase- and color-dependent variations. As a physical interpretation, the current phase-independent $CL(\lambda)$ would capture external dust absorption, while the additional $CL_\text{int}(\phi,\lambda)$ would account for potential intrinsic SN color  variations. This would be in line with recent findings of multiple SN color origins (e.g., \citealt{brout2021,nascimento2024}; Hand in prep.) \cite{dr2kenworthy} extends the SALT3 model to include an additional parameter, which captures this additional phase dependence of SN Ia color. This agrees with this work.

\subsubsection{Stretch dependence}
\label{sec:lcresx1dep}

The SN~Ia stretch is an intrinsic property that is directly connected to the  astrophysics of the SN~Ia explosion event \citep[e.g.,][and references therein]{scalzo2014,shen2021}. It has been shown to depend on the SN~Ia environment, and most likely, to depend on the progenitor age \citep[e.g.][]{hamuy1996,howell2007,nicolas2021}. Light-curve deviations as a function of SN~Ia stretch could provide interesting clues for a further understanding of SN~Ia physics, and it might help us to constrain the observed astrophysical biases in SN cosmology.

\begin{figure}
    \centering
    \includegraphics[width=\linewidth]{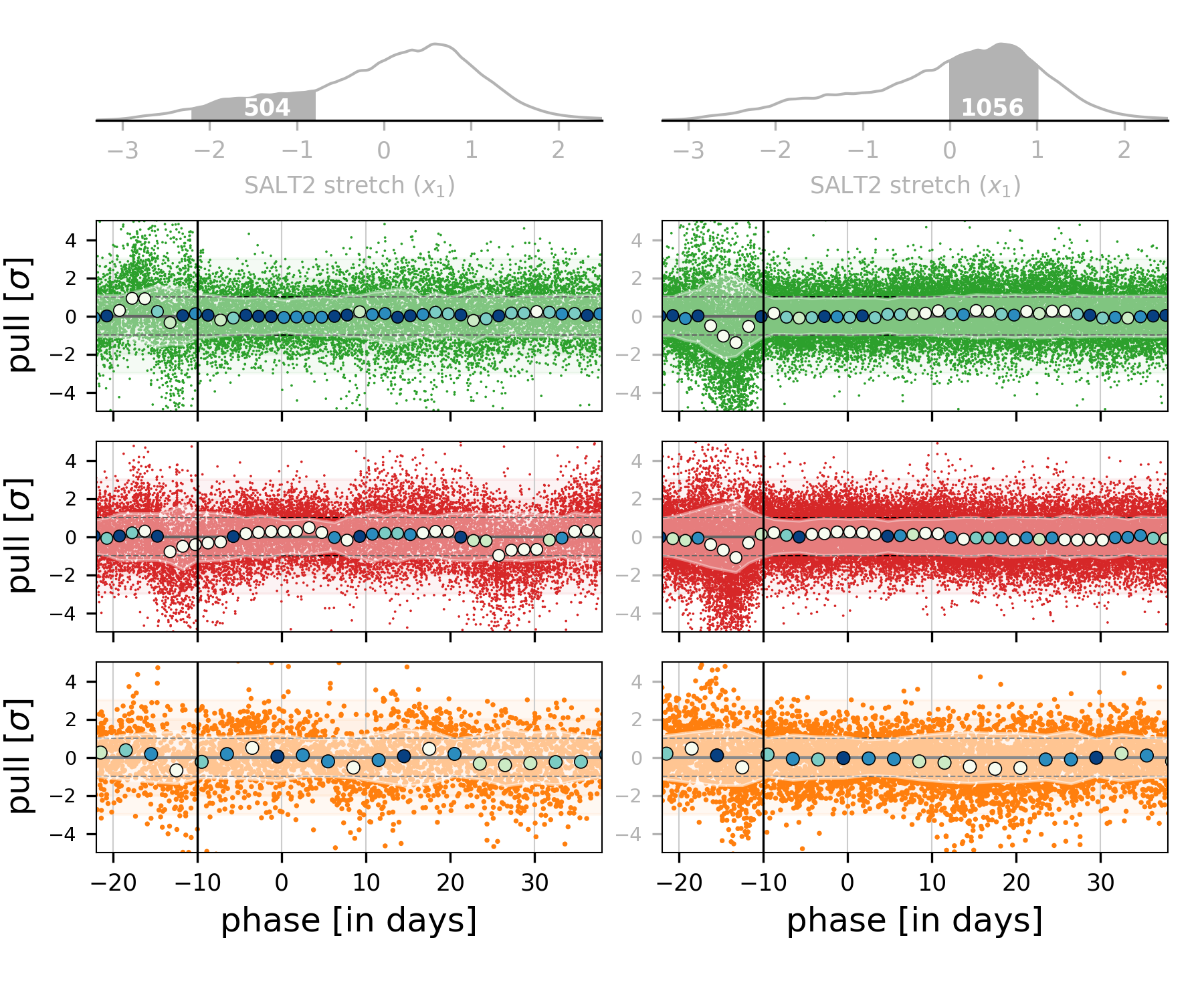} % .pdf available
    \caption{
    Similar to Fig.~\ref{fig:lcresidual_main}, but split per light-curve stretch ($x_1$) to sample both modes (\citealt{nicolas2021}, \citealt{dr2ginolina}). 
    For visibility, phase bins are 1.5~d in $g$ and $r$ and 3~d in $i$.
    \emph{Left side:} $x_1 \in [-2.2,-0.8]$ SNe~Ia. \emph{Right side:} $x_1 \in [0,+1.0]$ SNe~Ia. \emph{Top:} SN~Ia stretch distributions highlighting the sample considered on each side.
    }
\label{fig:lcresidual_x1var}
\end{figure}

Figure~\ref{fig:lcresidual_x1var} separately shows the SALT2 light-curve residuals for fast ($x_1\in[-2.2,-0.8]$) and slowly declining SNe~Ia ($x_1\in[+0,+1]$) (bins selected to sample the two stretch modes; see, e.g., \citealt{nicolas2021}). 
The light-curve residuals appear to be globally flat, suggesting that the stretch template correction ($M_1(\phi,\lambda)$) describes the data quantitatively well. 
This is particularly true for the high-stretch mode, which is the most populated mode. For these SNe, the sole significant deviation is in the $i$ band and is consistent with the effect we discussed in Sect.~\ref{sec:lcres} and showed in Fig.~\ref{fig:lcresidual_main} and in Fig.~\ref{fig:lcexample}: This means that the SALT2 template is smoother than the data. % and this may be fixed by a SALT2 retraining.
Low-stretch mode SNe~Ia, however, show some significant deviations in the $r$ band. The phase range near the second peak ($\phi \sim +25$~d) is particularly affected, suggesting that the $M_1(\phi,\lambda)$ structure itself would benefit from a retraining, and not just the SALT2 regularization terms. 
At maximum light, deviations are likely collateral issues caused by the late phase issue. The $g$ and $i$ bands are nonetheless correctly modeled, which suggests that this $r$-band issue does not significantly affect the overall fit. With the current SALT2 template, high-stretch mode SNe~Ia are better modeled than low-stretch ones overall.
This issue is likely due to the linear construction of the SALT model. Nonlinear models such as ParSNIP \citep{boone2021c} and SNooPy \citep{burns2011} are inherently more capable of simultaneously modeling both the extremes and the middle of the stretch distribution than SALT. Resolving the issue will require adding some degree of nonlinearity to the model or overcompensating with additional linear parameters.

Finally, the strong bump that is visible at low stretch in the $g$ band is a consequence of the poor modeling in SALT2 that we discussed in Sect.~\ref{sec:lcres}. It is not connected to the SN~Ia physics. It causes a dip in the residuals at high stretch because of nonphysical negative fluxes in the model and the opposite at low stretch.

\subsection{Goodness of fit as a function of host environment}
\label{sec:lcresvshost}

In this subsection, we investigate how SN~Ia light-curve residuals vary as a function of their environment. 
Deviations are expected because the SN~Ia parameters and, most importantly, the standardized SN~Ia magnitudes, have been shown to significantly depend on their host environments (see, e.g., \citealt{sullivan2010, roman2018, rigault2020, briday2022} and companion ZTF SN~Ia DR2 papers \citealt{dr2ginolina}, \citealt{dr2ginolinb}). 
The community is searching for an additional light-curve parameter that might capture these dependences. 

\begin{figure}
    \centering
    \includegraphics[width=\linewidth]{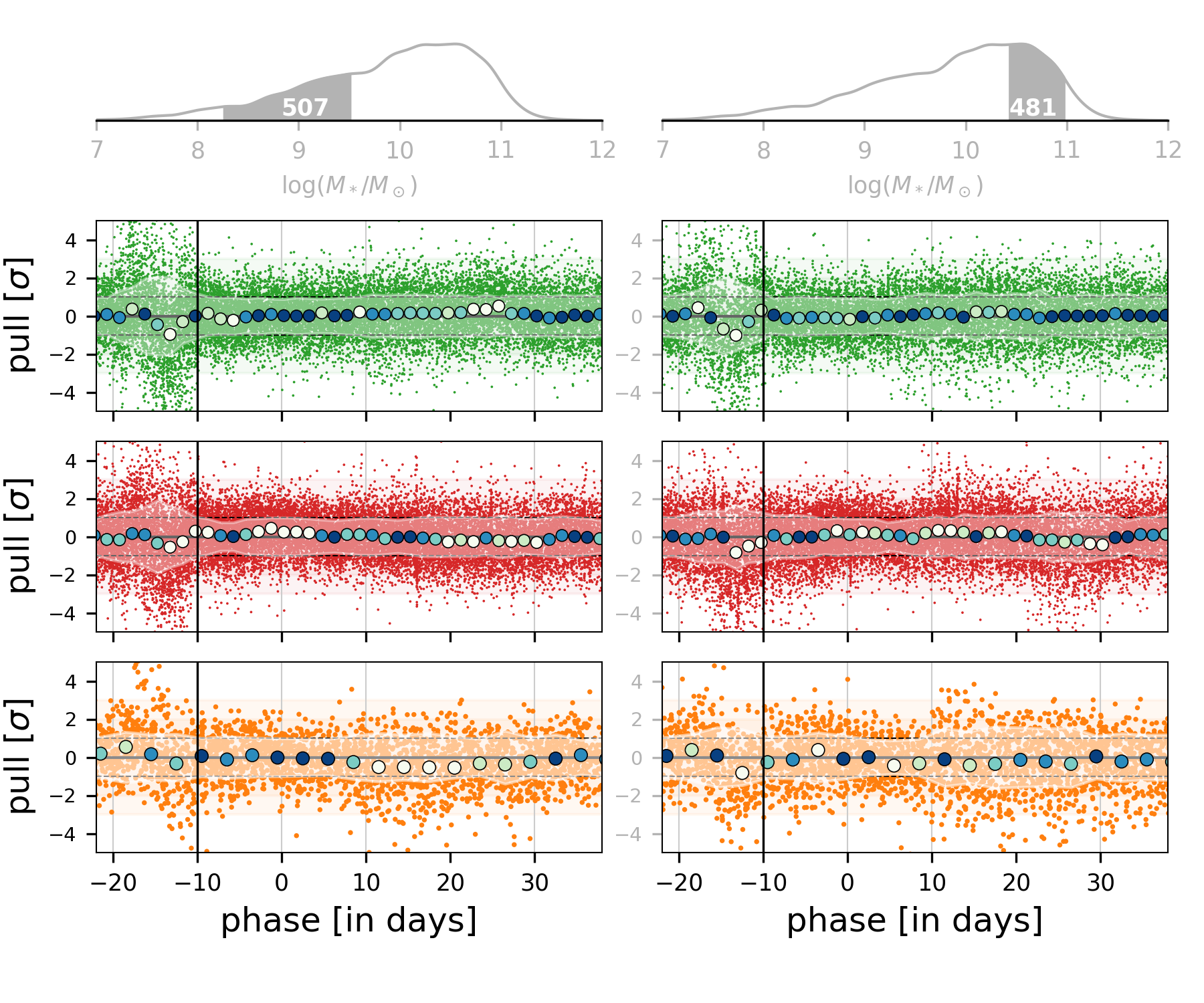} % .pdf available
    \includegraphics[width=\linewidth]{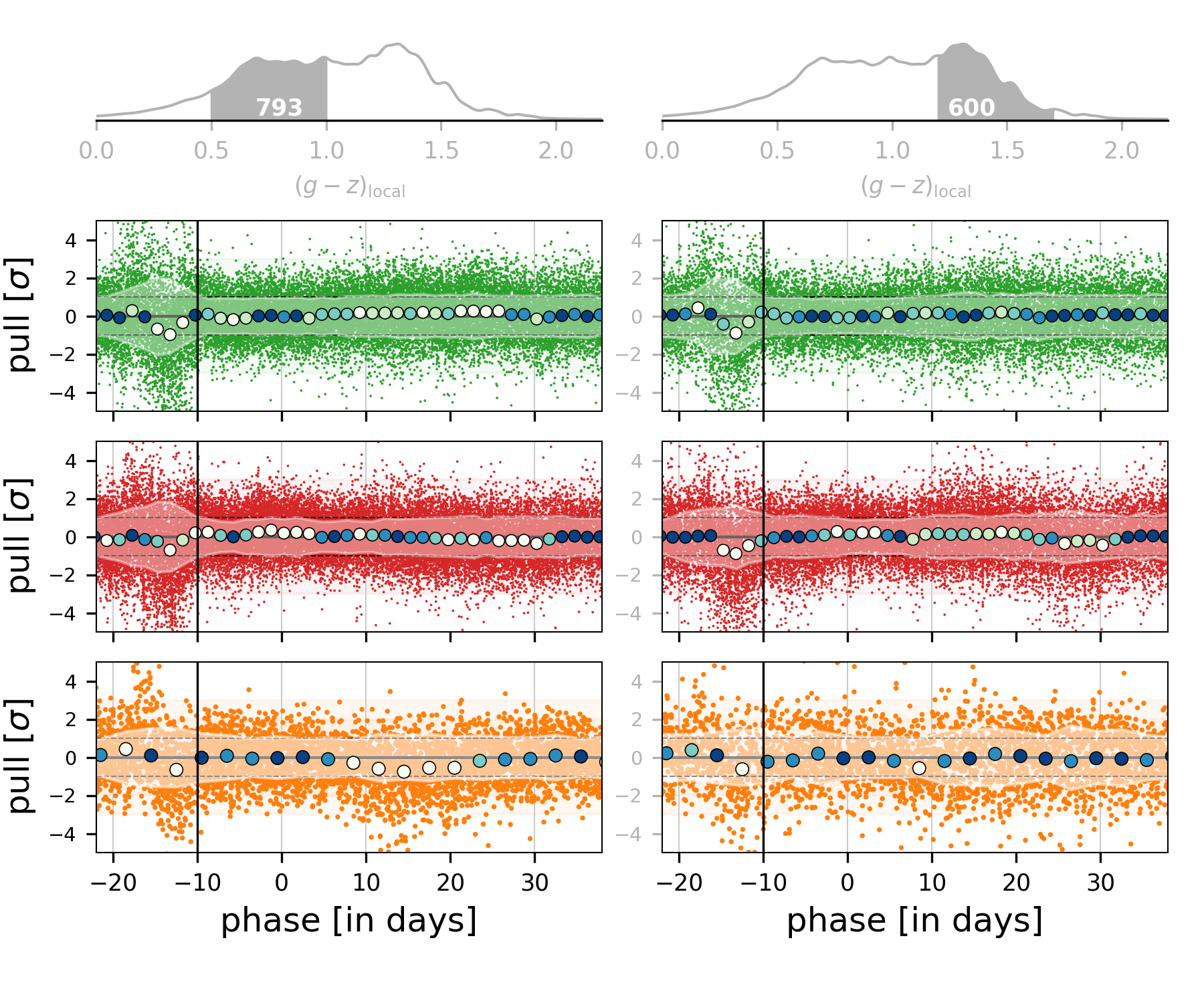} % .pdf available
    \caption{
    Similar to Fig.~\ref{fig:lcresidual_x1var} and \ref{fig:lcresidual_cvar}, but split according to global host mass ($\log(M_*/M_\odot)$, \emph{top}) or local color environment ($(g-z)_\text{local}$, \emph{bottom}). 
    }
\label{fig:lcresidual_hostvar}
\end{figure}

Following most literature analyses, we focused on the global host stellar mass ($\log(M_*/M_\odot)$) and the local environment color tracer ($(g-z)_\mathrm{local}$; see the parameter estimation in \cite{dr2rigault} and Smith et al. (in prep.). The light-curve residuals as a function of their environments are shown in Fig.~\ref{fig:lcresidual_hostvar}. Surprisingly, the light-curve residuals lack a significant host signature. Low- and high-mass host SNe~Ia have very similar residuals, as do locally blue- and red-environment targets. The only noticeable change is seen in the $i$ band at the time of the second peak, but this is very likely connected to the similar differences as a function of the SN light-curve stretch because low stretch-mode SNe~Ia only exist in massive host and/or red environments (see, e.g., \citealt{nicolas2021} and \cite{dr2ginolina}). This agrees with the results of \cite{jones2023}, who found limited evidence for a connection between light-curve residuals and host galaxy properties.

We therefore conclude that the light-curve parameter probably cannot capture SN-host dependences, at least not in the optical photometric bands probed by the ZTF. This conclusion might be revisited when the color- and phase-dependent variations presented in Sect.~\ref{sec:lcrescolordep} are accounted for.

\subsection{SALT2 versus SALT3}
\label{sec:testssalt}

In the main analysis, we used the SALT2 \citep{guy2010} algorithm trained by \cite{taylor2021}, which extends the original 2.4 version from \cite{betoule2014}. Recently, \cite{kenworthy2021} redesigned the training algorithm and created SALT3. In this section, we investigate whether our results hold when this SALT implementation is used. We used version 2.0 of SALT3, as available in sncosmo.

\begin{figure}
    \centering
    \includegraphics[width=\linewidth]{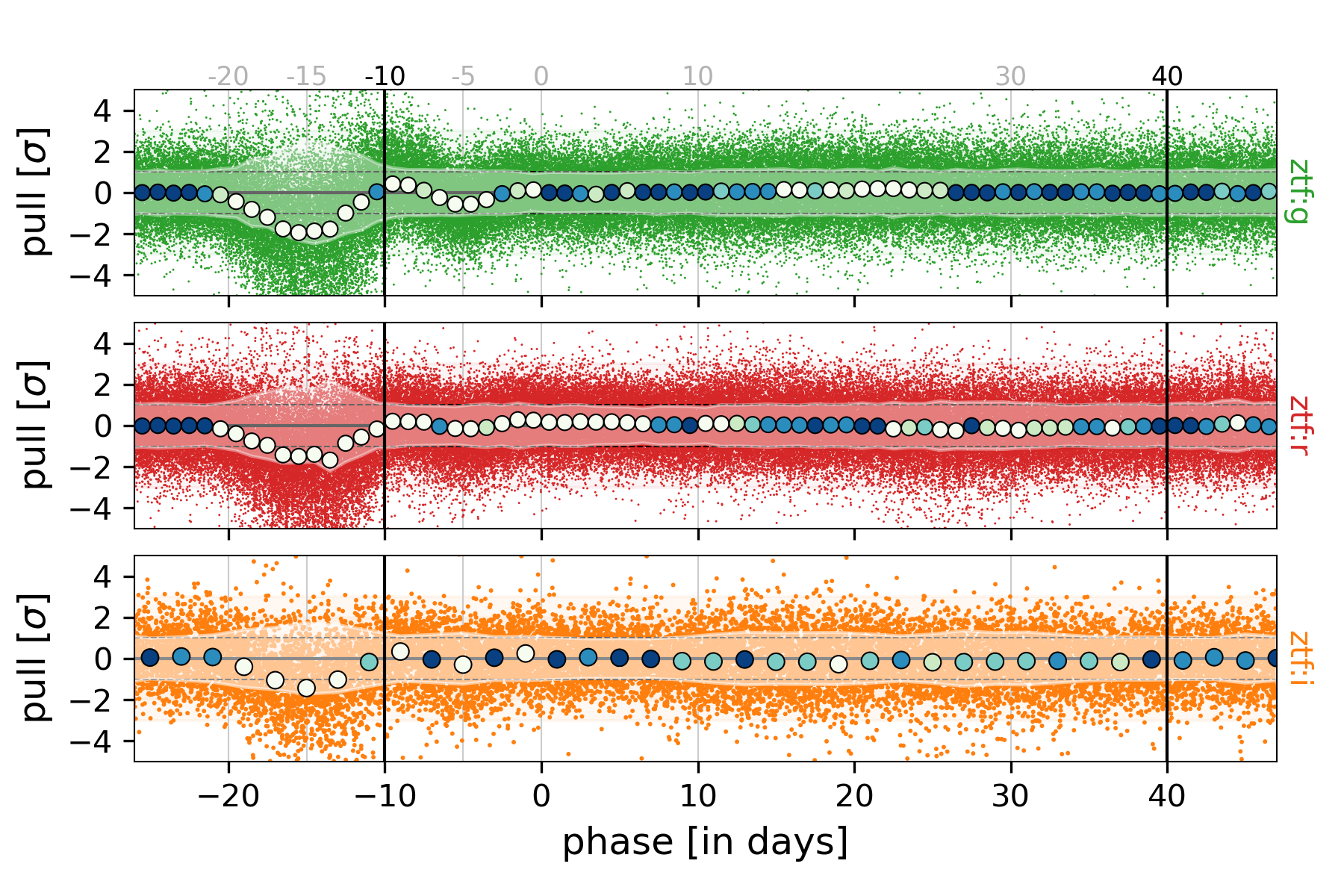} % .pdf available
    \caption{
    Similar as Fig.~\ref{fig:lcresidual_main}, but using the SALT3 light-curve fitting algorithm \citep{kenworthy2021}. The variations observed with SALT2 are still visible, notably pre maximum, where they are even stronger.
    }
\label{fig:lcresidual_salt3}
\end{figure}

The SALT3 light-curve fit residuals are shown in Fig.~\ref{fig:lcresidual_salt3}. The deviations observed with SALT2 that we showed in Fig.~\ref{fig:lcresidual_main} and discussed in  Sect.~\ref{sec:lcfitaccuracy} are still visible, and some are even stronger: 
\begin{enumerate}
    \item Early phase ($[-20, -10]$~d) fluxes are strongly overestimated by the light-curve fitter in all three bands.
    \item The model largely underestimates the $r$-band flux near peak brightness ($[-5, +5]$ d). 
    \item The second peak epoch ($[+15,+35]$ d) still deviates, but the deviations appear to be weaker in SALT3 than in SALT2. This is different from the two other issues, which are stronger in SALT3.
\end{enumerate}
In addition, there is a strong deviation at $-5$~d in SALT3 that does not seem significant in SALT2, especially in the $g$ band.

When investigating the aforementioned variations when splitting SNe~Ia by stretch (Sect.~\ref{sec:lcresx1dep}) or color (Sect.~\ref{sec:lcrescolordep}), we see very similar features and draw the same conclusions as we did when using SALT2. 
We thus conclude that the description of the ZTF light-curve data is very similar in SALT3 and SALT2. SALT3 seems to be slightly better at describing the second-peak epoch, but does significantly worse for pre-maximum light. This indicates that future training will probaby reduce the degree of regularization that needs to be applied.

\section{Variation in the light-curve parameter}
\label{sec:lcparam}

In this section, we investigate the dependence of the ZTF SN~Ia DR2 SALT2 parameters on the choices that are made to perform the light-curve fit, that is, in the $\phi\in[-10,+40]$~d phase range and for all three ZTF bands, when available. 
In Sect.~\ref{sec:phasevariations} we discuss the phase variations, and in Sect.~\ref{sec:gri_vs_gr} we investigate whether the $i$ band affects the derivation of the SALT2 light-curve parameters significantly.

%Investigating the impact of used phase-range, bands and cadence for non-ZTF samples it outside the scope of this paper. 

\subsection{Impact of a different fit phase range}
\label{sec:phasevariations}

\begin{figure}
    \centering
    \includegraphics[width=\linewidth]{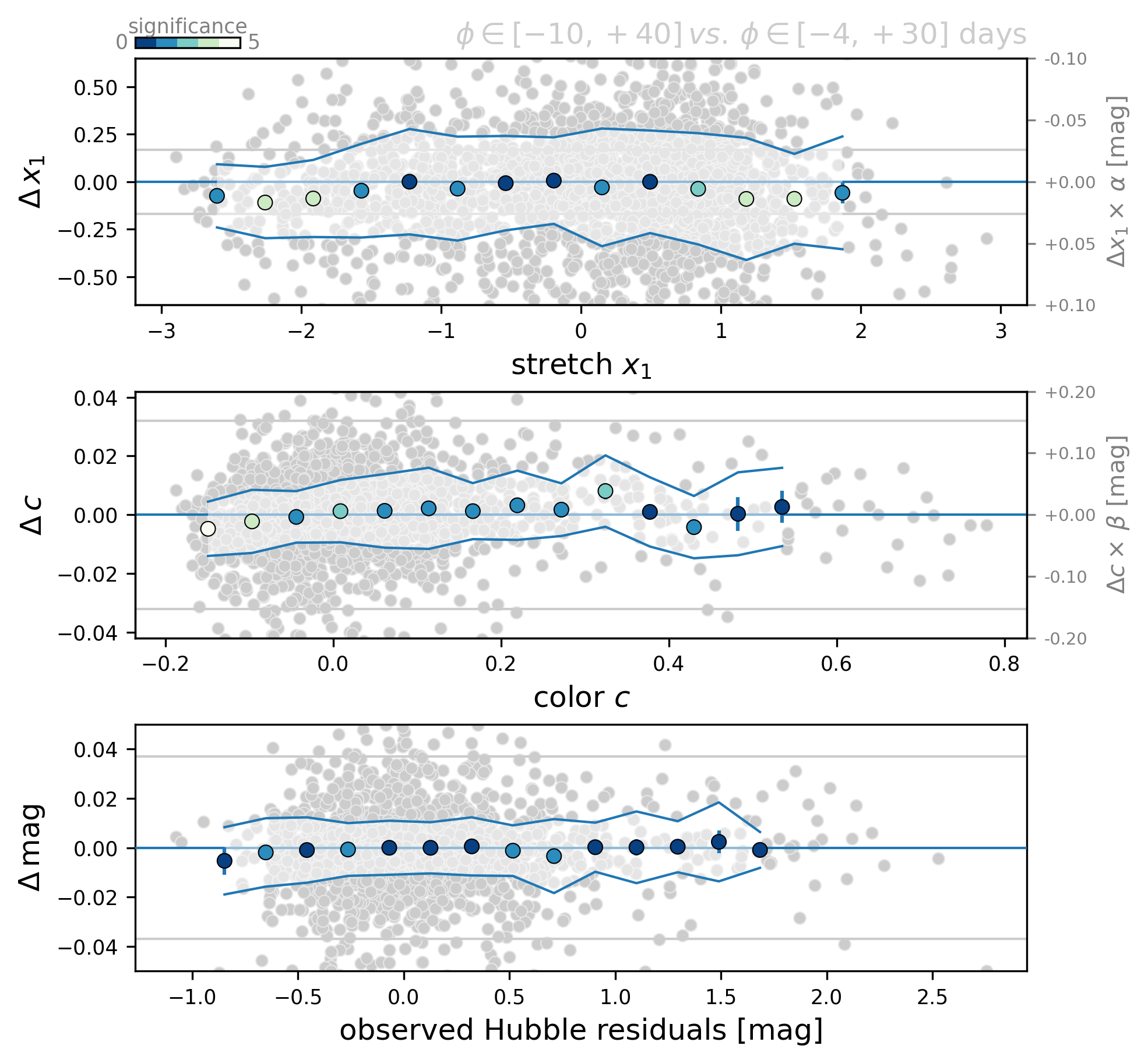} % .pdf available
    \caption{Impact of changing the fit phase range from 
    $\phi\in[-10,+40]$~d (${ref}$) to $\phi\in[-5,+30]$~d (${comp}$) on the estimate of the SALT2 light-curve parameters $x_1$ (\emph{top}), $c$ (\emph{middle}), and observed Hubble residuals, derived from $x_0$ assuming a default cosmology (\emph{bottom}). The x-axis in each panel shows the reference parameter (e.g., $c_{ref}$), and the y-axis shows the difference ($\Delta\,c \equiv c_{ref}-c_{comp}$). 
    In each panel, the horizontal gray lines show the median parameter error, and the marker color indicates the significance of the deviation from zero.
    The right ticks in the stretch and color panels show the corresponding magnitude impact, assuming $\alpha=-0.15$ and $\beta=3.15$.
    }
\label{fig:lcphase_variations}
\end{figure}

We compared the estimate of the SALT2 light-curve parameters when the reference phase range of $\phi\in[-10,+40]$~d was changed to either the smaller range ($\phi\in[-5,+30]$~d) or to the range with a longer baseline ($\phi\in[-20,+50]$~d). The comparison of the $\phi\in[-10,+40]$~d versus $\phi\in[-5,+30]$~d phase range is shown in Fig.~\ref{fig:lcphase_variations}. 

Fig.~\ref{fig:lcphase_variations} shows that the SALT2 parameters seem to be unbiased when the covered phase range is reduced. The scatter on $x_1$, the most sensitive parameter to phase coverage, is about the typical $x_1$ error ($\sim0.2$). The variations in SALT2 color ($c$) or peak magnitude (derived from $x_0$) are typically smaller than 30\% of the quoted errors. 
When the default phase-range is compared to the $\phi\in[-20,+50]$~d phase range instead of the shorter range $\phi\in[-5,+30]$~d (not shown), the conclusions are the same, but the variations are typically half of the variation presented in Fig.~\ref{fig:lcphase_variations}. 

We thus conclude that our SN light-curve parameter estimations are robust against reasonable changes in the fit phase range. This robustness is largely due to the impressive ZTF cadence (typically one photometric point per day; see \citealt{dr2rigault}).

\subsection{Missing the $i$ band}
\label{sec:gri_vs_gr}

We also investigated the impact of the use (or lack of use) of the $i$-band data when the light-curve parameters are investigated. As reported in \cite{dr2rigault} and Smith et al. (in prep.), only 46\% of our targets have at least one $i$-band detection within the $\phi\in[-10,+40]$~d phase range. Only 25\% have at least five detections. This means that it is questionable whether the SALT2 light-curve parameter estimation between SNe~Ia with and without $i$-band coverage is self-consistent. 

\begin{figure}
    \centering
    \includegraphics[width=\linewidth]{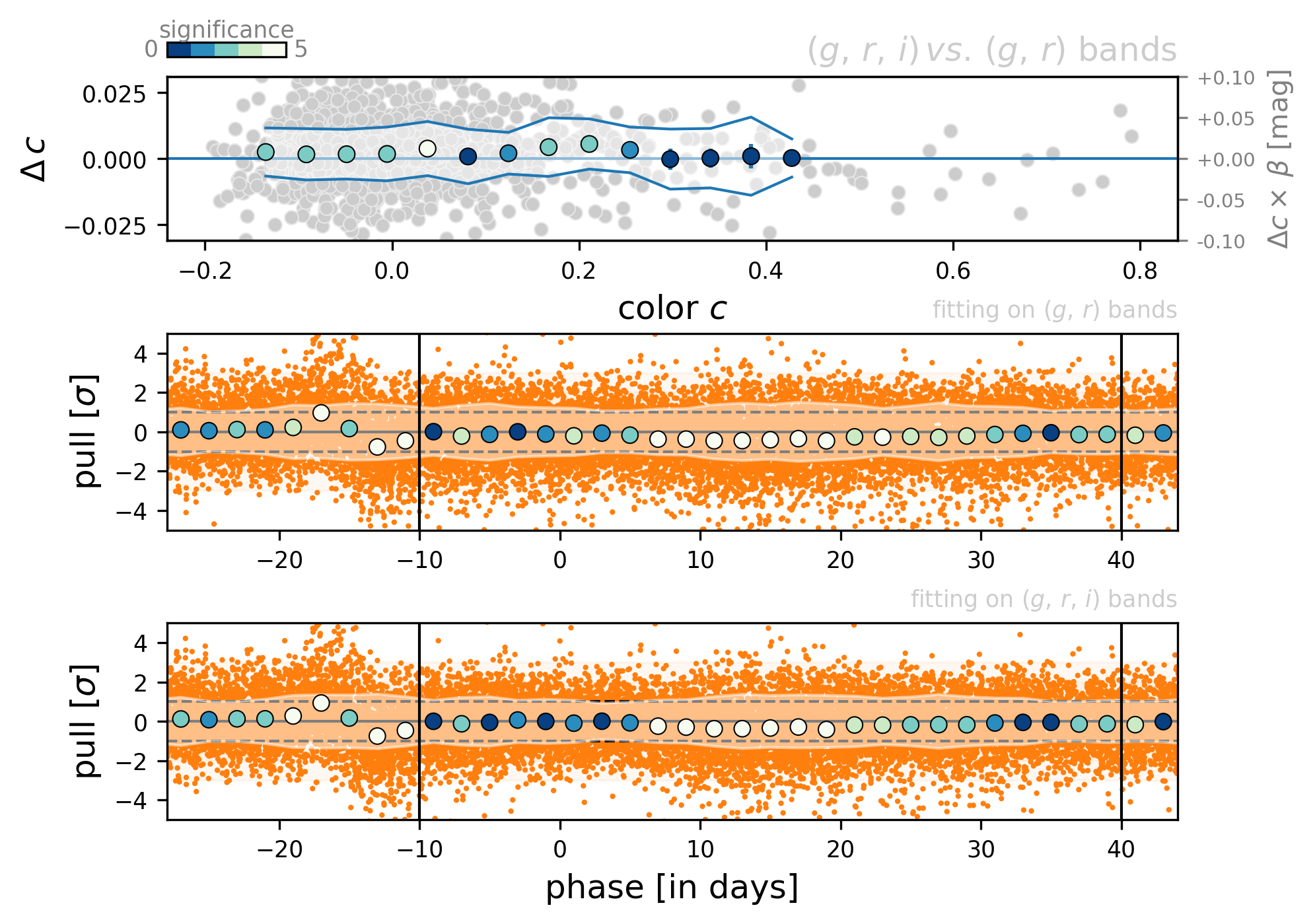} % .pdf available
    \caption{Impact of using $g$ and $r$ bands only to fit ZTF light-curve data with SALT2. The right ticks shows the conversion into magnitude assuming $\beta=3.15$. \emph{Top:} Variation in the color $c$ parameter. Cases without $i$-band data ($55\%$) are shown in gray.
    \emph{Middle:} $i$-band light-curve fit residuals (not used for the fit) shown following the layout of Fig.~\ref{fig:lcresidual_main}.
    \emph{Bottom:} For comparison, $i$-band light-curve fit residuals using all three bands, as shown in Fig.~\ref{fig:lcresidual_main}.
    }
\label{fig:lcgri_vs_gr}
\end{figure}

To test this, we compared the SALT2 parameters estimated with and without the $i$ band by refitting the light-curve parameters while discarding our $i$-band data. The resulting SALT2 color variations, which parameter is by far the most affected, and the resulting $i$-band light-curve residuals are shown in Fig.~\ref{fig:lcgri_vs_gr}.
This figure shows that the SALT2 colors vary by less than $0.01$, which is less than one-third of the typical SALT2 color parameter errors. This demonstrates that the SALT2 parameters estimated for targets with or without $i$-band data are compatible. 
This is further supported by the $i$-band light-curve residuals presented in Fig.~\ref{fig:lcgri_vs_gr}. The residual scatters are centered on zero and do not show more biases than were presented in this paper so far.  
The only small variation is the slight increase in scatter (running white band). This is to be expected because the $i$-band light curve never entered the fit, and the residuals we plot are the result of an extrapolation of the $g$, $r$ fit SALT2 model to the $i$ band. This remarkable agreement of the $i$-band light-curve residuals strengthens the quality of the ZTF SN~Ia DR2 band intercalibration.

% Why including the i-band
In light of the results presented in this subsection, the interest of acquiring $i$-band data for SN cosmology might be questioned. Here, we thus briefly summarize a few points.

We recall that this study is made in the context of the SALT2 algorithm, which only has one color parameter. Two bands are thus sufficient, in principle, and because the ZTF, $g$- and $r$-band observations are four times more frequent that the $i$-band observations ($\sim1.5$~d vs. $\sim5$~d cadence) and are deeper (20.5 mag vs. 20 mag; Smith et al., in prep), they largely overweight the $i$-band information during the SALT2 light-curve fit.
The $i$ band is important in SN cosmology for at least three reasons, however. 

% 1
The first reason is calibration: Three bands enable us to diagnose any calibration systematic within the bands because we can use two bands to predict the third band. 
% 2
The second reason is astrophysics: The existence of two colors in SN~Ia standardization is highly debated \citep[e.g.,][and references therein]{brout2021}. Three bands are thus needed to confirm or refute the existence of the two-color term in SN cosmology (see, e.g., Kenworthy et al. submitted.). If it is indeed confirmed, the $i$ band is moreover required to derive accurate distances (see also Sect.~\ref{sec:lcresvslcparam}). 
% 3
The third reason is operational: In red to near-infrared regions, SNe~Ia differ most from other SN types. The $i$ band is thus a powerful tool for distinguishing an SN~Ia from another type of transient. 

Cosmological surveys of SN then need to acquire data in at least three bands. 

\section{Conclusion}
\label{sec:conclusion}

We presented a detailed study of the ZTF DR2 SN~Ia light-curve residuals and compared them to the SALT2 model. We studied the accuracy of the model, which we did not retrain, to fit our $g$-, $r$-, and $i$-band dataset, which is made of 2625 SNe~Ia. In particular, we investigated (1) the phase range that should be considered for the light-curve fit, (2) the effect that is created by the lack of $i$-band data for 55\% of our targets in the estimation of their light-curve parameters, and (3) whether the SALT2 model describes SN~Ia light curves equally well, regardless of their parameter or host environments. 
This study is relevant for SN cosmology because the use of the SALT2 light-curve fitter is a core ingredient for the derivation of distances. An inaccuracy of the model that explains the data could then lead to biases in the derivation of the cosmological parameters. 

Our conclusions are listed below.
\begin{enumerate}
    \item Globally, the current implementation of SALT2 (2.4 or T21) represents our dataset remarkably well, even though our sample is much larger than the sample used for the model training. In all three optical ZTF bands, the model residuals from rest-frame phases $\phi$ between $-10$ to $+40$~d systematically deviate from zero by only a tenth of the errors.
    
    \item The light-curve flux error estimates of the ZTF SN~Ia DR2 are compatible with the observed data when a 2.5\%, 3.5\% and 6\% SN flux scatter is included to account for noise that is introduced by the difference-image pipeline. The amplitude of this error floor is compatible with findings based on simulations reported in the companion paper (\cite{dr2amenouche}).
    
    \item The SALT2 model is not able to describe the early SN~Ia light-curve phases ($\phi<-10$~d), where it systematically overestimates the observed flux. This is particularly true for $g$ and $r$ bands. This issue is most likely due to a lack of training data in this phase range and should be fixed by including datasets such as the ZTF in the training set. 
    
    \item When we compared the SNe~Ia per host environment, we saw no sign of light-curve residual differences. This strongly suggests that the SN optical light curves do not contain a signature that would enable us to absorb the observed  environmental dependences in SN standardized magnitude (also known as the steps). An improved early epoch modeling and color (see below) may change this conclusion.
    
    \item We showed many small but significant biases (nonzero residuals) in many epochs of the light curves, even in the $\phi\in[-10,+40]$~d phase range. Most notably, the $r$-band flux at maximum light is underestimated, and many deviations are visible around the second peak in all three bands. The SALT2 model seems to be too smooth globally in comparison to what the data would favor. It is very likely that a retraining of the SALT2 model with additional data would absorb these variations.
    
    \item We reported significant light-curve residual variations for blue and red SNe~Ia that probably cannot be absorbed by a retraining the SALT2 model. They rather strongly suggest the existence of a third component in the form of a phase-dependent color term. Its impact on the overall light-curve modeling might be small, but its estimation could be important for cosmology because claims of multiple sources of SN color are often made.
    
    \item The light-curve variations are stronger in low-stretch SNe~Ia, which is again expected because these are rarer and fainter and are thus not as well represented in the training dataset.
    
    \item The SALT2 parameters reported in the ZTF SN~Ia DR2 are robust against phase variations ($\phi \in [-5, +30]$ or $\phi \in [-20, +50]$~d in place of the default $\phi \in [-10, +40]$~d) or against the absence of $i$-band data.
    
    \item Using SALT3 (version 2.0) in place of SALT2 (T21 or 2.4-JLA) leads to very similar conclusions. SALT3 nonetheless shows more light-curve residual deviations at maximum light, but a clear improvement at second-peak phases.
    
\end{enumerate}

Based on the results of this paper, we strongly advocate that SN cosmological analyses should include a retraining of the light-curve fitter model to accurately represent the complexity of the considered dataset. This is critical for samples that are significantly larger than the sample that is used to train the model. Consequently, cosmological analyses based on ZTF data will require a retraining of the SALT model.
We finally encourage modellers to consider searching for a color-dependent term that might explain the observed color-dependence variations in the light-curve residuals.

\begin{acknowledgements}
  Based on observations obtained with the Samuel Oschin Telescope 48-inch and the 60-inch Telescope at the Palomar Observatory as part of the \textit{Zwicky} Transient Facility project. ZTF is supported by the National Science Foundation under Grants No. AST-1440341 and AST-2034437 and a collaboration including current partners Caltech, IPAC, the Weizmann Institute of Science, the Oskar Klein Center at Stockholm University, the University of Maryland, Deutsches Elektronen-Synchrotron and Humboldt University, the TANGO Consortium of Taiwan, the University of Wisconsin at Milwaukee, Trinity College Dublin, Lawrence Livermore National Laboratories, IN2P3, University of Warwick, Ruhr University Bochum, Northwestern University and former partners the University of Washington, Los Alamos National Laboratories, and Lawrence Berkeley National Laboratories. Operations are conducted by COO, IPAC, and UW.
  SED Machine is based upon work supported by the National Science Foundation under Grant No. 1106171
  The ZTF forced-photometry service was funded under the Heising-Simons Foundation grant \#12540303 (PI: Graham).
  This project has received funding from the European Research Council (ERC) under the European Union's Horizon 2020 research and innovation program (grant agreement n 759194 - USNAC). This project is supported by the H2020 European Research Council grant no. 758638. This work has been supported by the Agence Nationale de la Recherche of the French government through the program ANR-21-CE31-0016-03.
L.G. acknowledges financial support from the Spanish Ministerio de Ciencia e Innovaci\'on (MCIN) and the Agencia Estatal de Investigaci\'on (AEI) 10.13039/501100011033 under the PID2020-115253GA-I00 HOSTFLOWS project, from Centro Superior de Investigaciones Cient\'ificas (CSIC) under the PIE project 20215AT016 and the program Unidad de Excelencia Mar\'ia de Maeztu CEX2020-001058-M, and from the Departament de Recerca i Universitats de la Generalitat de Catalunya through the 2021-SGR-01270 grant. Y.-L.K. has received funding from the Science and Technology Facilities Council [grant number ST/V000713/1]. This work has been enabled by support from the research project grant ‘Understanding the Dynamic Universe’ funded by the Knut and Alice Wallenberg Foundation under Dnr KAW 2018.0067. SD acknowledges support from the Marie Curie Individual Fellowship under grant ID 890695 and a Junior Research Fellowship at Lucy Cavendish College. We thank the Heising-Simons Foundation for supporting the research program of SRK.

\end{acknowledgements}


\begin{thebibliography}{}

  % A
\bibitem[Aldering et al.(2002)]{aldering2002} Aldering, G., Adam, G., Antilogus, P., et al.\ 2002, \procspie, 4836, 61.

\bibitem[Amenouche et al.(2024)]{dr2amenouche} Amenouche, M., Smith,
  M., Rosnet, P., et al.\ 2024,
  arXiv:2409.04650. doi:10.48550/arXiv.2409.04650, (ZTFSI)

% B
\bibitem[Barbary et al. (2023)]{sncosmo2023} Barbary et al. 2023, Zenodo. doi: 10.5281/zenodo.8393360.

\bibitem[Bellm et al.(2019)]{bellm2019} Bellm, E.~C., Kulkarni, S.~R., Graham, M.~J., et al.\ 2019, \pasp, 131, 018002.

\bibitem[Betoule et al.(2014)]{betoule2014} Betoule, M., Kessler, R., Guy, J., et al.\ 2014, \aap, 568, A22.

\bibitem[Boone et al.(2021a)]{boone2021a} Boone, K., Aldering, G., Antilogus, P., et al.\ 2021, \apj, 912, 70.

\bibitem[Boone et al.(2021b)]{boone2021b} Boone, K., Aldering, G., Antilogus, P., et al.\ 2021, \apj, 912, 71.

\bibitem[Boone(2021)]{boone2021c} Boone, K.\ 2021, \aj, 162, 275.

\bibitem[Briday et al.(2022)]{briday2022} Briday, M., Rigault, M., Graziani, R., et al.\ 2022, \aap, 657, A22.

\bibitem[Brout \& Scolnic(2021)]{brout2021} Brout, D. \& Scolnic, D.\ 2021, \apj, 909, 26.

\bibitem[Brout et al.(2022)]{brout2022} Brout, D., Scolnic, D., Popovic, B., et al.\ 2022, \apj, 938, 110.

\bibitem[Burns et al.(2011)]{burns2011} Burns, C.~R., Stritzinger, M., Phillips, M.~M., et al.\ 2011, \aj, 141, 19.


% C
% D
\bibitem[Dhawan et al.(2015)]{dhawan2015} Dhawan, S., Leibundgut, B., Spyromilio, J., et al.\ 2015, \mnras, 448, 1345.

\bibitem[Dimitriadis et al.(2019)]{dimitriadis2019} Dimitriadis, G., Foley, R.~J., Rest, A., et al.\ 2019, \apjl, 870, L1.

\bibitem[Deckers et al.(2022)]{deckers2022} Deckers, M., Maguire, K., Magee, M.~R., et al.\ 2022, \mnras, 512, 1317.

\bibitem[Deckers et al.(2023)]{deckers2023} Deckers, M., Graur, O., Maguire, K., et al.\ 2023, \mnras, 521, 4414.
% E

% F
\bibitem[Fakhouri et al.(2015)]{fakhouri2015} Fakhouri, H.~K., Boone, K., Aldering, G., et al.\ 2015, \apj, 815, 58.

\bibitem[Firth et al.(2015)]{firth2015} Firth, R.~E., Sullivan, M., Gal-Yam, A., et al.\ 2015, \mnras, 446, 3895.

\bibitem[Folatelli et al.(2010)]{folatelli2010} Folatelli, G., Phillips, M.~M., Burns, C.~R., et al.\ 2010, \aj, 139, 120.

\bibitem[Freedman et al.(2019)]{freedman2019} Freedman, W.~L., Madore, B.~F., Hatt, D., et al.\ 2019, \apj, 882, 34.

  % G
\bibitem[Ginolin et al.(2024,a)]{dr2ginolina} Ginolin, M., Rigault, M., Smith, M., et al.\ 2024, arXiv:2405.20965. doi:10.48550/arXiv.2405.20965, (ZTFSI)

\bibitem[Ginolin et al.(2024,b)]{dr2ginolinb} Ginolin, M., Rigault, M., Copin, Y., et al.\ 2024, arXiv:2406.02072. doi:10.48550/arXiv.2406.02072, (ZTFSI)


\bibitem[Goobar \& Leibundgut(2011)]{goobar2011} Goobar, A. \& Leibundgut, B.\ 2011, Annual Review of Nuclear and Particle Science, 61, 251.

\bibitem[Grayling et al.(2024)]{grayling2024} Grayling, M., Thorp, S., Mandel, K.~S., et al.\ 2024, \mnras, 531, 953.
  
\bibitem[Graham et al.(2019)]{graham2019} Graham, M.~J., Kulkarni, S.~R., Bellm, E.~C., et al.\ 2019, \pasp, 131, 078001.

\bibitem[Guy et al.(2007)]{guy2007} Guy, J., Astier, P., Baumont, S., et al.\ 2007, \aap, 466, 11.

\bibitem[Guy et al.(2010)]{guy2010} Guy, J., Sullivan, M., Conley, A., et al.\ 2010, \aap, 523, A7.


% H
\bibitem[Hamuy et al.(1996)]{hamuy1996} Hamuy, M., Phillips, M.~M., Suntzeff, N.~B., et al.\ 1996, \aj, 112, 2438.

\bibitem[Howell et al.(2007)]{howell2007} Howell, D.~A., Sullivan, M., Conley, A., et al.\ 2007, \apjl, 667, L37.

% I

% J
\bibitem[Jha et al.(2007)]{jha2007} Jha, S., Riess, A.~G., \& Kirshner, R.~P.\ 2007, \apj, 659, 122.

\bibitem[Jones et al.(2023)]{jones2023} Jones, D.~O., Kenworthy, W.~D., Dai, M., et al.\ 2023, \apj, 951, 22.

% K
\bibitem[Kasen(2006)]{kasen2006} Kasen, D.\ 2006, \apj, 649, 939.

\bibitem[Kelly et al.(2010)]{kelly2010} Kelly, P.~L., Hicken, M., Burke, D.~L., et al.\ 2010, \apj, 715, 743.

\bibitem[Kelsey et al.(2023)]{kelsey2023} Kelsey, L., Sullivan, M., Wiseman, P., et al.\ 2023, \mnras, 519, 3046.

\bibitem[Kenworthy et al.(2021)]{kenworthy2021} Kenworthy, W.~D., Jones, D.~O., Dai, M., et al.\ 2021, \apj, 923, 265.

\bibitem[Kenworthy et al.(2024)]{dr2kenworthy} Kenworthy, W.~D., Goobar, A., Jones, D.~O. et al.\ 2024, \aap, submitteb (ZTF SI)

% L
\bibitem[L{\'e}get et al.(2020)]{leget2020} L{\'e}get, P.-F., Gangler, E., Mondon, F., et al.\ 2020, \aap, 636, A46.

% M
\bibitem[Mandel et al.(2022)]{mandel2022} Mandel, K.~S., Thorp, S., Narayan, G., et al.\ 2022, \mnras, 510, 3939.

\bibitem[M{\"u}ller-Bravo et al.(2022)]{mullerbravo2022} M{\"u}ller-Bravo, T.~E., Sullivan, M., Smith, M., et al.\ 2022, \mnras, 512, 3266.


% N
\bibitem[Nascimento et al.(2024)]{nascimento2024} Nascimento, C.~S., Fran{\c{c}}a, J.~P.~C., \& Reis, R.~R.~R.\ 2024, Astronomy and Computing, 49, 100866.

\bibitem[Nicolas et al.(2021)]{nicolas2021} Nicolas, N., Rigault, M., Copin, Y., et al.\ 2021, \aap, 649, A74.

% O
% P
\bibitem[Papadogiannakis et al.(2019)]{papadogiannakis2019} Papadogiannakis, S., Goobar, A., Amanullah, R., et al.\ 2019, \mnras, 483, 5045.

\bibitem[Perlmutter et al.(1999)]{perlmutter1999} Perlmutter, S., Aldering, G., Goldhaber, G., et al.\ 1999, \apj, 517, 565.

\bibitem[Pessi et al.(2022)]{pessi2022} Pessi, P.~J., Hsiao, E.~Y., Folatelli, G., et al.\ 2022, \mnras, 510, 4929.

\bibitem[Phillips(1993)]{phillips1993} Phillips, M.~M.\ 1993, \apjl, 413, L105.

\bibitem[Planck Collaboration et al.(2020)]{planck2020} Planck Collaboration, Aghanim, N., Akrami, Y., et al.\ 2020, \aap, 641, A6.

\bibitem[Popovic et al.(2021)]{popovic2021} Popovic, B., Brout, D., Kessler, R., et al.\ 2021, \apj, 913, 49. 

% Q
% R
\bibitem[Roman et al.(2018)]{roman2018} Roman, M., Hardin, D., Betoule, M., et al.\ 2018, \aap, 615, A68.

\bibitem[Rose et al.(2021)]{rose2021} Rose, B.~M., Baltay, C., Hounsell, R., et al.\ 2021, arXiv:2111.03081. doi:10.48550/arXiv.2111.03081

\bibitem[Riess et al.(1996)]{riess1996} Riess, A.~G., Press, W.~H., \& Kirshner, R.~P.\ 1996, \apj, 473, 88.

\bibitem[Riess et al.(1998)]{riess1998} Riess, A.~G., Filippenko, A.~V., Challis, P., et al.\ 1998, \aj, 116, 1009.

\bibitem[Riess et al.(2022)]{riess2022} Riess, A.~G., Yuan, W., Macri, L.~M., et al.\ 2022, \apjl, 934, L7.

\bibitem[Rigault et al.(2013)]{rigault2013} Rigault, M., Copin, Y., Aldering, G., et al.\ 2013, \aap, 560, A66.

\bibitem[Rigault et al.(2015)]{rigault2015} Rigault, M., Aldering, G., Kowalski, M., et al.\ 2015, \apj, 802, 20.

\bibitem[Rigault et al.(2020)]{rigault2020} Rigault, M., Brinnel, V., Aldering, G., et al.\ 2020, \aap, 644, A176.

\bibitem[Rigault et al.(2024)]{dr2rigault} Rigault, M., Smith, M., Goobar, A., et al.\ 2024, arXiv:2409.04346. doi:10.48550/arXiv.2409.04346, (ZTFSI)

  
\bibitem[Rubin et al.(2022)]{rubin2022} Rubin, D., Aldering, G., Astraatmadja, T.~L., et al.\ 2022, arXiv:2206.10632.

% S
\bibitem[Saunders et al.(2018)]{saunders2018} Saunders, C., Aldering, G., Antilogus, P., et al.\ 2018, \apj, 869, 167.

\bibitem[Scalzo et al.(2014)]{scalzo2014} Scalzo, R., Aldering, G., Antilogus, P., et al.\ 2014, \mnras, 440, 1498.

\bibitem[Schlafly et al.(2011)]{Schlafly2011} Schlafly, E. F., Finkbeiner, D. P.\ 2011, \apj, 737, 103.

\bibitem[Scolnic et al.(2018)]{scolnic2018} Scolnic, D.~M., Jones, D.~O., Rest, A., et al.\ 2018, \apj, 859, 101.

\bibitem[Shen et al.(2021)]{shen2021} Shen, K.~J., Blondin, S., Kasen, D., et al.\ 2021, \apjl, 909, L18.

\bibitem[Sullivan et al.(2010)]{sullivan2010} Sullivan, M., Conley, A., Howell, D.~A., et al.\ 2010, \mnras, 406, 782.

% T
\bibitem[Taylor et al.(2021)]{taylor2021} Taylor, G., Lidman, C., Tucker, B.~E., et al.\ 2021, \mnras, 504, 4111.

\bibitem[Taylor et al.(2023)]{taylor2023} Taylor, G., Jones, D.~O., Popovic, B., et al.\ 2023, \mnras, 520, 5209.

\bibitem[Tripp(1998)]{tripp1998} Tripp, R.\ 1998, \aap, 331, 815

% U
% V
% W
\bibitem[Wiseman et al.(2022)]{wiseman2022} Wiseman, P., Vincenzi, M., Sullivan, M., et al.\ 2022, \mnras, 515, 4587.

% X
% Y
% Z
\end{thebibliography}
\end{document}